\documentclass[11pt]{article}

\usepackage{subcaption}
\usepackage[dvipsnames]{xcolor}
\usepackage[totalheight = 21cm, totalwidth = 15cm]{geometry}
\usepackage{amssymb,amsmath,amsfonts,amsbsy,graphicx,ctable,colortbl}
\usepackage{bm,color}
\usepackage{cite}
\usepackage[%
             colorlinks=true,urlcolor=MidnightBlue,linkcolor=MidnightBlue,citecolor=OliveGreen,
             pdfpagelabels=true,hypertexnames=true,
            plainpages=false,naturalnames=false,
             ]{hyperref}

\usepackage{epsfig}

\newcommand{\calR}{{\cal R}}

\newcommand{\ltsima}{$\; \buildrel < \over \sim \;$}
\newcommand{\lsim}{\lower.5ex\hbox{\ltsima}}
\newcommand{\gtsima}{$\; \buildrel > \over \sim \;$}
\newcommand{\gsim}{\lower.5ex\hbox{\gtsima}}

\newcommand{\be}{\begin{equation}}
\newcommand{\ee}{\end{equation}}
\newcommand{\bea}{\begin{eqnarray}}
\newcommand{\eea}{\end{eqnarray}}

\begin{document}


\begin{center}


{\Large \bf Cross-tests of CMB features in the primordial spectra}

\vskip 0.5cm

{
Spyros Sypsas
}

\vskip 0.5cm

{\it
Departamento de F\'isica, FCFM, Universidad de Chile \\ Blanco Enclada 2008, Santiago, Chile
}

\vskip 1cm

\end{center}

\begin{abstract}

The recent {\sc Planck} data on the power spectrum of temperature anisotropies of the cosmic microwave background marginally support deviations from the $\Lambda$CDM model at several multipoles. With a view towards current and forthcoming observational surveys, we trace these features to other observables like the scalar bispectrum and the tensor power spectrum. A  possible detection of such bumps in these channels would increase their statistical significance shedding light on the ultra violet mechanisms responsible for their appearance in the data.

\end{abstract}



\section{Introduction}
Inflation~\cite{inflation} is currently the main paradigm of early universe dynamics compatible with the current data which point towards a Gaussian and nearly scale invariant power spectrum of the cosmic microwave background (CMB) temperature anisotropies~\cite{Ade:2015xua}. Since the proposal of the original idea in the 80's there has been a vast variety of models realising it, many of which draw inspiration from ultra violet (UV) completions of the QFT and GR frameworks like string theory and supergravity. 

An indispensable characteristic of these UV realisations is the presence of new high energy degrees of freedom, whose interaction with inflation may leave imprints on the CMB observables. Like in particle physics, where new particles appear as bumps over the standard model background, such imprints would be manifested in small deviations from the standard cosmological model. Indeed, this is exactly what the {\sc Planck} satelite has shown~\cite{Aghanim:2015xee} reconfirming earlier observations from {\sc Wmap}~\cite{Hinshaw:2012aka}, a fact which has sparked an intense effort to study these features~\cite{jointanalysis}.

The difference in the cosmological setup is of course that data are very hard to collect resulting in a low statistical significance of at most $2\sigma$ of these bumps~\cite{dev-scaleinv}. However, as in particle physics, a way to increase the significance of deviations from the known model is to look for them in different channels, or in the cosmological language, in different CMB spectra.

The {\sc Planck} mission is currently constraining the scalar bispectrum, while several ongoing and future surveys are looking for primordial tensor modes. In the meantime, a large amount of data is expected in the next few years from large scale structure observations.
In anticipation of the coming data, we workout how these bumps in the power spectrum, if real, would manifest themselves in the scalar bispectrum and the tensor power spectrum.
\section{Method}
We are interested in computing how the presence of sharp features, which are produced by a brief deviation of the inflaton from the slow-roll dynamics\footnote{Note that inflation is not interrupted by this mechanism. One can easily realise a setup where $\epsilon$ remains small but its derivative, encoded in $\eta$, gets large -- $\mathcal{O}(0.5)$ -- during a short time interval. This is enough to produce features in the power spectrum.}, propagate from one spectrum to another. Let us thus define two spectra as $S_1$ and $S_2$ and a set of parameters $\epsilon_i=\{\epsilon,\eta,c_s\}$, where $\epsilon=\dot H/H^2$, with $H$ denoting the Hubble scale, $\eta=\dot\epsilon/\epsilon H$ and $c_s$ the soundspeed of the scalar perturbations. The correlators $S_i$ could be any among the scalar, tensor power spectra or higher point functions. 

The effective field theory of inflation allows us to write down a model independent Lagrangian describing the dynamics of curvature perturbations around the characteristic energy scale of inflation, the Hubble constant $H$, which may be used to compute the aforementioned correlators.
Within this context, the idea is to setup an equation that connects the scalar power spectrum to other spectra.

The relation is set up as follows:
\begin{itemize}
\item[$\star$] One splits the parameters of the Lagrangian into a slow-roll background and a fast contribution. Then using the \emph{in-in} formalism~\cite{Maldacena:2002vr} 
one computes the corrections on the slow-roll part of $S_1$ induced by the fast-rolling parts of the theory. At this stage one has equations of the form 
\be \label{Pofe}
\Delta S_1 = \int_0^\infty dt \; {\cal D}^t_1(\epsilon_i) \; \calR^n ,
\ee
where $\Delta S_1$ is the correction on the spectrum $S_1$, ${\cal D}_1^t$ is a diferrencial operator with respect to time, while $n\in \mathbb{N}$, is a number counting the order of the correlator $S_1$.
\item[$\star$] One then exploits the fact that the de Sitter mode functions $\calR \propto e^{ikt}$ contain the Fourier measure. Using a trick to define parameters in the time interval $[-\infty,\infty]$, one may Fourier invert Eq.~\eqref{Pofe}, ending up with formulas of the following type:
\be \label{eofP}
\epsilon_i = \int_{-\infty}^\infty dk \; {\cal D}_2^k \; \Delta S_1(k),
\ee
that is, the parameters $\epsilon_i$ are written as functions of the spectrum.
\item[$\star$] One then computes another correlator, say $S_2$, to obtain another equation similar to \eqref{Pofe}:
\be \label{Bofe}
\Delta S_2 = \int_{-\infty}^\infty dt \;{\cal D}_3^t(\epsilon_i) \;\calR^m .
\ee
Finally, one substitutes the $\epsilon_i$'s from Eq.~\eqref{eofP} obtaining formulas of the form
\be \label{BofS}
\Delta S_1 =  {\cal D}_4^k[\Delta S_2].
\ee
With such a formula at hand, one may compute how features in the spectrum $S_2$ propagate to the correlator $S_1$. 
\end{itemize}

\section{Results}
\subsection{Scalar Power/Bispectrum Correlation} \label{sec:sta}
\begin{figure}[h]
 \begin{center}
 \begin{tabular}{cc} 
  \includegraphics[width=.4\textwidth]{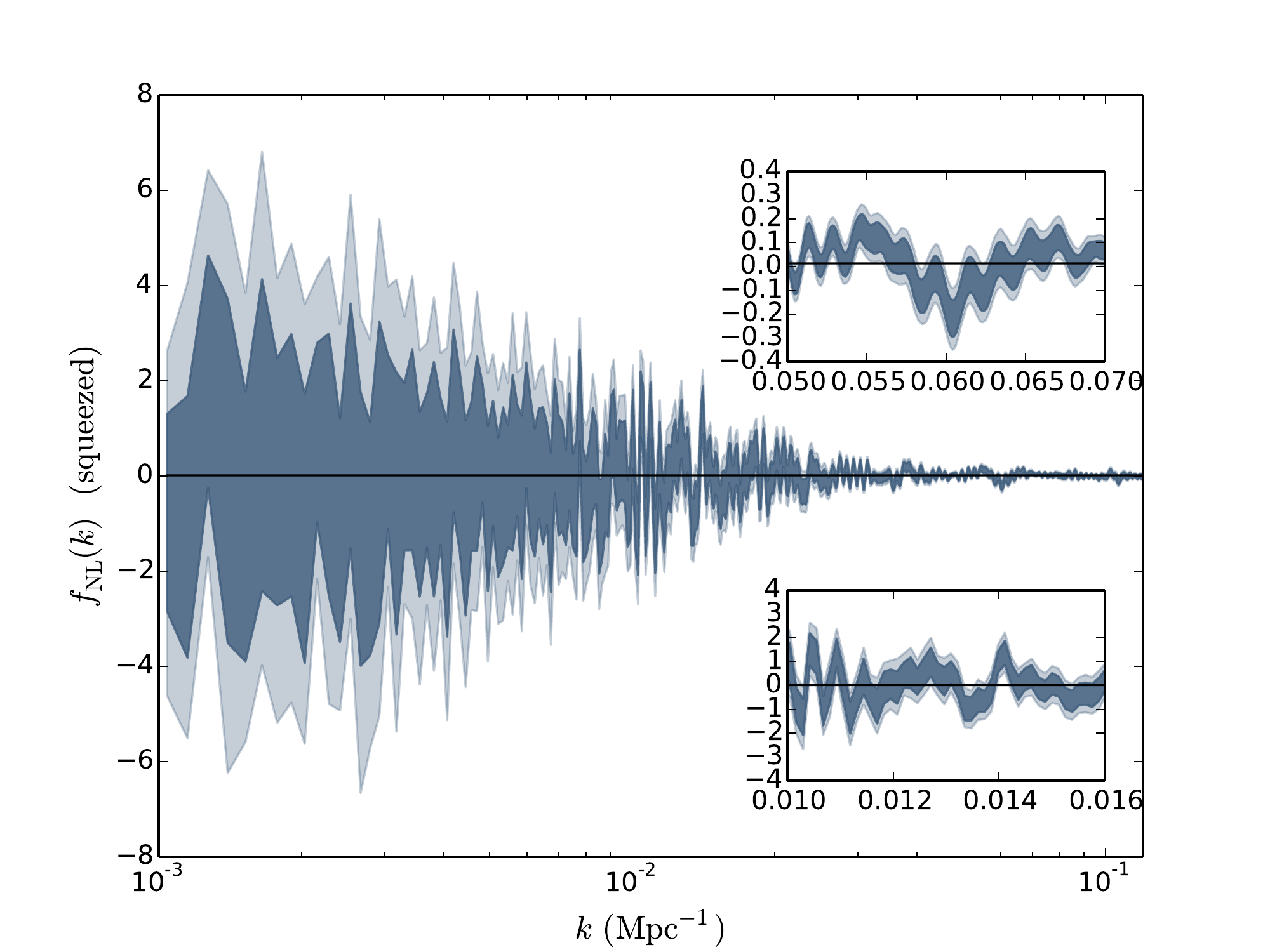} & \includegraphics[width=.4\textwidth]{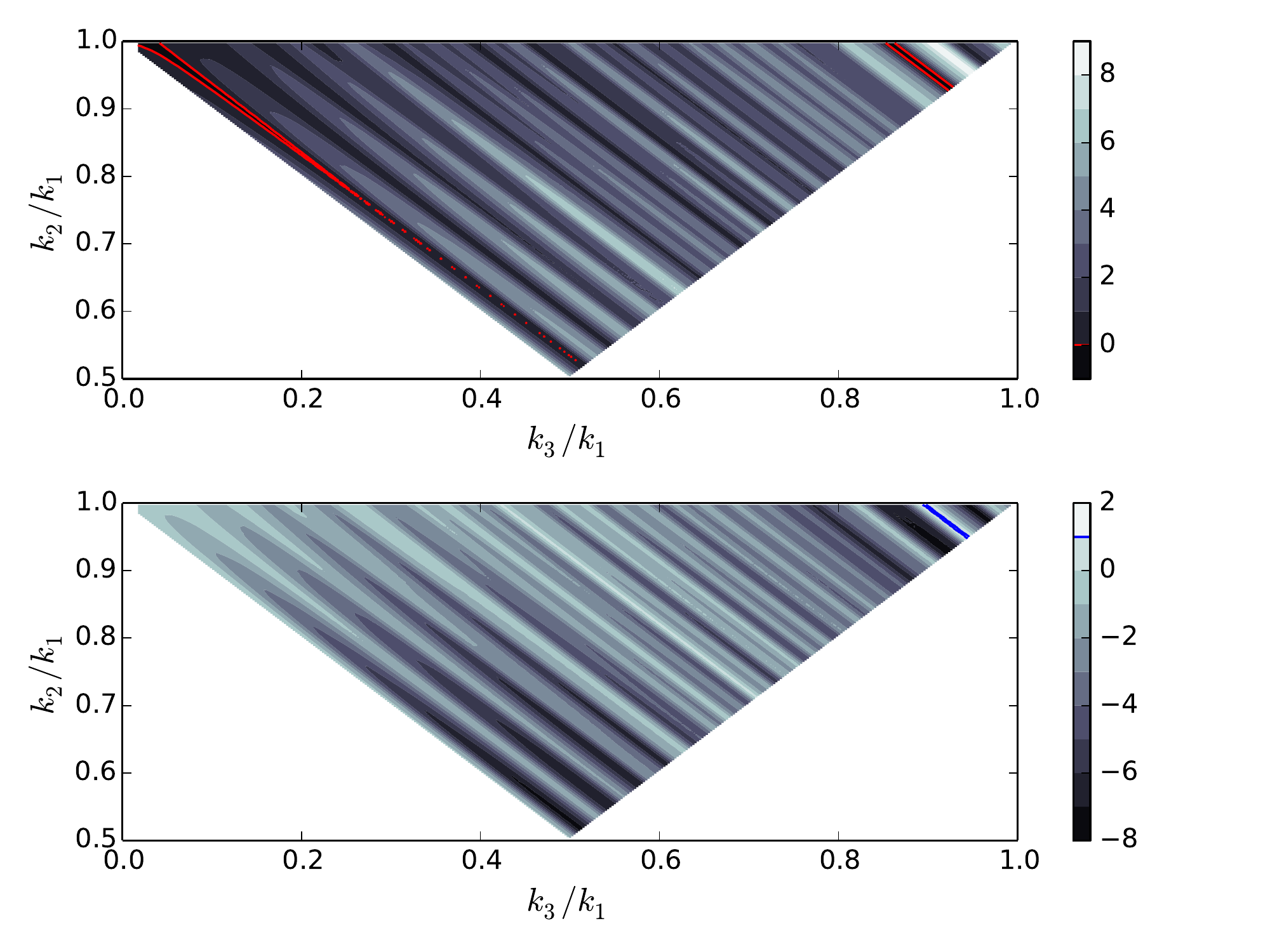}
 \end{tabular}
 \end{center}
 \caption{{\it Left panel}: bispectrum amplitude at the squeezed limit, $k_3\sim0$. The feature at $k=0.06$ Mpc$^{-1}$ corresponds to the $\ell=800$, $2\sigma$ feature of the angular power spectrum. \\ {\it Right panel}: difference between featureless and featureful bispectrum amplitudes for the scale $k_1=0.06$ Mpc$^{-1}$. Red and blue contours indicate the $\vec{k}$-regions where this difference is nonzero by more than 2$\sigma$.}
 \label{fig:1}
\end{figure}

Here, based on Ref.~\cite{Appleby:2015bpw}, we apply the aforementioned algorithm to the case where $S_2=P_\calR $ is the power spectrum and $S_1=B_\calR$, the scalar bispectrum\footnote{In this work, we applied a different method in order to correlate the two spectra based on the generalised slow-roll formalism~\cite{Stewart:2001cd}. However, the exact same formula has been obtained in~\cite{Palma:2014hra} using the Fourier inversion. Due to lack of spacetime, we have chosen to comment only on the latter, which we have also used to obtain the rest of the results presented here.}. We assume that features are present only in the Hubble parameters, while $c_s$ is kept constant\footnote{If features are present in both parameters, one can not invert Eq.~\eqref{Pofe} to get the relation~\eqref{eofP}. However, this is possible if one assumes a relation between the two quantities~\cite{Mooij:2015cxa}.}. The result is:
\be \label{BofS1}
B_\calR(k_1,k_2,k_3) \!\!  \propto \!\! 
\left[ \left( 1+x^2+y^2 \right) \frac{x + y + xy}{16}  + \frac{x^2 + y^2 + (xy)^2}{8}  - \frac{xy}{8} \right] (1-n_\calR) + \frac{xy}{8}\alpha_\calR,  
\ee
where $k_1=k,\;k_2=xk,\;k_3=yk$ and
%
\be 
1-n_\calR = d_{\log k}\log P_\calR \;,\qquad \alpha_\calR = d_{\log k}^2\log P_\calR,
\ee
are the spectral index and its running, respectively. 
Before applying it to real data, one may first test the above formula. We did that using a typical model with a step in the inflationary potential known to produce features. We computed numerically the left hand side and the right hand side (RHS) of Eq.~\eqref{BofS1} and found excelent agreement. Next, we may input the {\sc Planck} power spectrum data into the RHS of our formula~\eqref{BofS1} and get a prediction of the scalar bispectrum in the case of features in the Hubble flow parameters. We present the results in Fig.~\ref{fig:1}.

\subsection{Scalar/Tensor Power Spectrum Correlation}
Next, based on Ref.~\cite{Palma:2016wqu}, we apply the procedure to the case where $S_2=P_S$ is the scalar power spectrum and $S_1=P_T$ is the tensor one. The result is:
%
\be \label{TofP}
\Delta P_T = - 6 \iint d\ln k \;\; \epsilon  \Delta P_S,
\ee
We see that features in the tensor spectrum are suppressed due to slow-roll, which was known before~\cite{Hu:2014hoa}, but also because of the integral structure of Eq.~\eqref{TofP}, which smooths out any feature of $P_S$. This indicates that, the tensor power spectrum remains practically scale invariant even if the scalar spectrum admits scale dependent features. 

Note that this statement does not categorically exclude local enhancements -- i.e. features -- in the tensor spectrum\footnote{For example, in models with non Bunch-Davies vacuum, one can obtain observable features in $P_T$~\cite{Broy:2016zik}.}. It just says that under fairly general assumptions the features appearing in the {\sc Planck} data, e.g. the $\ell=20$ dip in the angular power spectrum, will not be observable in the tensor spectrum. Reversing the statement, the very interesting case of the detection of a feature in $P_T$ at multipole $\ell=20$, would either point to an astrophysical origin of the bump, or would be an indicator of a very exotic inflationary model. 

\subsection{Features in the Bispectrum}
Finally, based on Ref.~\cite{Gong:2017vve}, we want to see how features appear in the bispectrum if a correlation between spectra is not possible because e.g. all parameters admit a brief period of fast changes. Here, $S_1$ and $S_2$ are the bispectrum shape function evaluated at two mode configurations with momenta $k_1=k,\;k_2=x_1k,\;k_3=y_1k$ and $p_1=k,\;p_2=x_2k,\;p_3=y_2k$. This leads us to two main results:
\subsubsection{Bispectrum Consistency Relation for Features}
If the background parameters experienced brief deviations from slow-roll during inflation then the bispectrum should obey the following relation:
\begin{align}
\label{eq:con-rel1}
S_\calR(k,x,y) & = \Big[ \alpha_1 S_1(x,y) - \alpha_2 S_2(x,y) \Big] S_\calR \left( \frac{1+x+y}{1+x_1+y_1}k,x_1,y_1 \right) 
\nonumber\\
& \quad - \Big[ \beta_1 S_1(x,y) - \beta_2 S_2(x,y) \Big] S_\calR \left( \frac{1+x+y}{1+x_2+y_2}k,x_2,y_2 \right) \, ,
\end{align}
where $\alpha_{1,2}$ and $\beta_{1,2}$ are parameters depending on $x_{1,2}$ and $y_{1,2}$ and the partial shapes $S_{1,2}$ can be found in~\cite{Gong:2017vve}.
For example, fixing equilateral $(x_1,y_1)=(1,1)$ and flat $(x_2,y_2)=(1/2,1/2)$ shapes, we obtain
\begin{align}
\label{eq:eq-flat}
S_\calR(k,x,y) & = \frac{18(x+y+xy) - 15(1+x^2+y^2)}{(1+x+y)^2} S_\calR^{} \left(\frac{1+x+y}{3}k,1,1\right)
\nonumber\\
& \quad -16 \frac{x+y+xy - (1+x^2+y^2)}{(1+x+y)^2} S_\calR^{} \left(\frac{1+x+y}{2}k,\frac{1}{2},\frac{1}{2}\right) \, ,
\end{align}
implying that the amplitude of the bispectrum at any point $(k,x,y)$ should be related to the corresponding amplitudes evaluated at $\left(\frac{1+x+y}{3}k,1,1\right)$ and $\left(\frac{1+x+y}{2}k,1/2,1/2\right)$.
This can be viewed as a consistency relation since it holds only in the case of features. We have tested the above formula numerically using a simple step model as in Sec.~\ref{sec:sta}.

\subsubsection{Featured Bispectrum Templates}
Motivated by the form of the consistency relation of Eq.~\eqref{eq:con-rel1}, we may construct templates for the featured bispectrum. We thus replace the amplitudes on the RHS of Eq.~\eqref{eq:con-rel1} with a sine function which is a typical choise to model oscillating features in the spectra~\cite{Ade:2015ava}. We have\footnote{The same template can be written with log arguments in the sine functions, which is another typical profile modeling features~\cite{Ade:2015ava}.}:
\begin{equation}
S_{\alpha\beta}(k,x,y) = S_{\alpha_1\alpha_2}(x,y) \sin \Big[ (1+x+y)\omega_1 k+\phi \Big] + S_{\beta_1\beta_2}(x,y) \sin \Big[ (1+x+y)\omega_2 k+\phi \Big],
\end{equation}
where the partial modulating shapes, including the known -- equilateral, orthogonal, flattened -- templates, can be found in~\cite{Gong:2017vve}.

Interestingly, the template contains two frequencies $\omega_1$ and $\omega_2$ stemming from the factors $\frac{1}{1+x_1+y_1}$ and $\frac{1}{1+x_2+y_2}$ in the amplitudes involved in the consistency relation~\eqref{eq:con-rel1}. In our construction it is evident that oscillating features may involve as many frequencies as couplings in the cubic Lagrangian. This is in agreement with the {\sc Planck} results which favour such a multifrequency distribution for features~\cite{Ade:2015ava}.
\section{Concluding Remarks}
If the deviations from the $\Lambda$CDM line observed by {\sc Planck} and {\sc Wmap} are real indicators of new physics then they should also show up in other spectra. In these works we have traced how features appear in different observables. We found that {\it i}) the bispectrum should be enhanced around the scale $k=0.06$ Mpc$^{-1}$ if during inflation the Hubble parameters experience short deviations from slow-roll without interupting inflation; {\it ii}) the tensor power spectrum remains scale invariance unless some exotic mechanism produces the features; {\it iii}) in the case where all background parameters admit nontrivial time dependence, the bispectrum obeys certain consistency relations. In the case of oscillating spectra, there can be as much frequencies involved as cubic couplings in the Lagrangian, while all the known templates may appear as modulating shapes.

\section*{Acknowledgements}
We have reported on work done in collaboration with Gonzalo Palma, Jinn-Ouk Gong, Arman Shafieloo, Dhiraj Hazra, Stephen Appleby, Basti\'an Pradenas and W\'alter Riquelme, all of whom we wish to thank. SS is supported by the FONDECYT grant N$^{\rm o}$ 3160299.

\end{document}